\documentclass[twocolumn,superscriptaddress,amsmath,amssymb,prb,aps,floatfix]{revtex4-1}
\usepackage{graphicx,color}
\usepackage{epstopdf}
\pdfoutput=1

\newcommand{\tcb}{\textcolor{blue}}

\begin{document}

\title{Step-edge assisted large scale FeSe monolayer growth on epitaxial Bi$_2$Se$_3$ thin films}

\author{J. Fik\'a\v{c}ek}
\affiliation{Institute of Physics, Academy of Sciences of the Czech Republic, Na Slovance 2, CZ-18221 Prague 8, Czech Republic}
\affiliation{Department of Condensed Matter Physics, Faculty of Mathematics and Physics, Charles University, Ke Karlovu 5, CZ-12116 Prague 2, Czech Republic}


\author{M. Vondr\'a\v{c}ek}
\affiliation{Institute of Physics, Academy of Sciences of the Czech Republic, Na Slovance 2, CZ-18221 Prague 8, Czech Republic}

\author{V. Stetsovych}
\affiliation{Institute of Physics, Academy of Sciences of the Czech Republic, Na Slovance 2, CZ-18221 Prague 8, Czech Republic}

\author{P. Proch\'azka}
\affiliation{CEITEC - Central European Institute of Technology, Brno University of Technology, Purkyňova 123, 612 00 Brno, CZ}

\author{S. Pr\r u\v{s}a}  
\affiliation{CEITEC - Central European Institute of Technology, Brno University of Technology, Purkyňova 123, 612 00 Brno, CZ}

\author{L. Kormo\v{s}}
\affiliation{CEITEC - Central European Institute of Technology, Brno University of Technology, Purkyňova 123, 612 00 Brno, CZ} 

\author{J. \v{C}echal}
\affiliation{CEITEC - Central European Institute of Technology, Brno University of Technology, Purkyňova 123, 612 00 Brno, CZ}

\author{O. Caha}
\affiliation{Institute of Physics, Academy of Sciences of the Czech Republic, Na Slovance 2, CZ-18221 Prague 8, Czech Republic}
\affiliation{CEITEC - Central European Institute of Technology, Brno University of Technology, Purkyňova 123, 612 00 Brno, CZ}
\affiliation{Department of Condensed Matter Physics, Masaryk University, Kotlarska 2, CZ-61137 Brno}

\author{T. Sk\'ala}
\affiliation{Department of Condensed Matter Physics, Faculty of Mathematics and Physics, Charles University, Ke Karlovu 5, CZ-12116 Prague 2, Czech Republic}

\author{P. Vlaic}
\affiliation{Biophysics Department, University of Medicine and Pharmacy “Iuliu Hatieganu”, 400023 Cluj-Napoca, Romania}

\author{K. Carva}
\affiliation{Department of Condensed Matter Physics, Faculty of Mathematics and Physics, Charles University, Ke Karlovu 5, CZ-12116 Prague 2, Czech Republic}

%


\author{G. Springholz}
\affiliation{Institute of Semiconductor and Solid State Physics, Johannes Kepler University, Altenbergerstrasse 69, A-4040 Linz, Austria}


\author{J. Honolka}
\affiliation{Institute of Physics, Academy of Sciences of the Czech Republic, Na Slovance 2, CZ-18221 Prague 8, Czech Republic}
\email{honolka@fzu.cz}

\date{\today}


\begin{abstract}
The interest in Fe-chalcogenide unconventional superconductors is intense after the critical temperature of FeSe was reported enhanced by more than one order of magnitude in the monolayer limit at the interface to an insulating oxide substrate. In heterostructures comprising interfaces of FeSe with topological insulators, additional interesting physical phenomena is predicted to arise e.g. in form of {\it topological superconductivity}.
So far superconductive properties of Fe-chalcogenide monolayers were mostly studied by local scanning tunneling spectroscopy experiments, which can detect pseudo-gaps in the density of states as an indicator for Cooper pairing. Direct macroscopic transport properties which can prove or falsify a true superconducting phase were rarely reported due to the difficulty to grow films with homogeneous material properties. Here we report on a promising growth method to fabricate continuous carpets of monolayer thick FeSe on molecular beam epitaxy grown Bi$_2$Se$_3$ topological insulator thin films. In contrast to previous works using atomically flat cleaved bulk Bi$_2$Se$_3$ crystal surfaces we observe a strong influence of the high step-edge density (terrace width about 10~nm) on MBE-grown Bi$_2$Se$_3$ substrates, which significantly promotes the growth of coalescing FeSe domains with small tetragonal crystal distortion without compromising the underlying Bi$_2$Se$_3$ crystal structure.

\end{abstract}

\pacs{}

\maketitle


\section{Introduction \label{sec:intro}}
FeSe is one of the youngest representatives of Fe-based superconductors with a critical temperature ($T_{\text c}$) of 8~K in the bulk. Due to its simple two-element tetragonal PbO structure with a layer unit cell (UC) height $c = 5.52$~\AA ~it is since then studied intensively in search of the physics behind unconventional superconductivity (SC).\\ 
In the last years the interest in Fe-chalcogenides grew again after $T_{\text c}$ of FeSe was found enhanced by a factor $\times 20$ in the thin-film limit at interfaces to non-conductive oxides (also coined {\it interfacial superconductivity}). First theoretical approaches address the influence of strain~\cite{Tan2013}, electron doping~\cite{Miyata2015}, and band bending effects at the interface, in particular the competition between spin density waves and antiferromagnetism~\cite{Bazhirov2013}, closely related to the appearance of unconventional SC. For an overview on recent developments see e.g. Ref.[~\cite{Liu2015}].\\
Interfacing Fe-chalcogenides with the special material class of topological insulators (TIs)~\cite{Hasan2010} yet promises further interesting physics as the effect of {\it topological superconductivity} is predicted to appear~\cite{RevModPhys.83.1057}.
In 2016 prototypical chalcogenide-based TI single crystals Bi$_2$Ch$_3$ (Ch = Se, Te) were used as substrates to grow FeCh ontop (FeSe/Bi$_2$Se$_3$ and FeTe/Bi$_2$Te$_3$, respectively)~\cite{Cavallin2016, Eich2016}. FeCh layers were grown by a simple two-step procedure consisting of (i) initial room temperature deposition of Fe atoms onto {\it in-situ} cleaved Bi$_2$Ch$_3$ crystals and (ii) subsequent annealing at $T = 300^{\circ}$C. This leads to the formation of nanometer wide FeCh islands of variable thicknesses 1-5~UC with areal coverages well below 50\%. Local scanning tunneling spectroscopy (STS) on FeSe islands on Bi$_2$Se$_3$(FeSe is SC in the bulk) showed no gap in the local density of states (LDOS) in the accessible range of temperatures $T\geq 5$K\cite{Eich2016}, while surprisingly respective FeTe on Bi$_2$Te$_3$(FeTe is normal conducting in the bulk) develops a pseudo-gap in the LDOS at 6.5~K~\cite{He2013}. It is important to stress here that the detection of a pseudo-gap by STS generally is not sufficient to prove SC.\\ 
For a full understanding of SC properties in the FeCh monolayer limit, non-local properties like lateral transport need to be studied, which requires homogeneous and continuous FeCh layers.
Scanning tunneling microscopy (STM) studies~\cite{Cavallin2016, Eich2016} so far focused on selected local FeCh islands and thereby widely neglected the reported discontinuous surface morphology on a larger scale with areas whose surface termination and stoichiometry are yet unknown.
In fact, from simple stoichiometry arguments, FeCh growth from Fe atoms on intact Bi$_2$Ch$_3$ surface should lead to formation of Bi-rich phases. \\
Here we report on an alternative approach based on molecular beam epitaxy (MBE) Bi$_2$Se$_3$ thin films, where step-edge assisted growth allows to fabricate well-ordered $\mu$m-scale wide FeSe monolayer carpets at an intact topological insulator interface. FeSe carpets consist of three equally prominent 60$^{\circ}$ rotated FeSe domains resolved in low-energy electron microscopy (LEEM) with a lateral size of 30-50~nm, well beyond the expected Cooper pair length scales of 1~nm for Fe-chalcogenides~\cite{Singh2013}. Combining standard STM techniques with non-local, but highly surface sensitive ultraviolet photomemission spectroscopy (UPS) and low-energy ion scattering (LEIS) we show that FeSe growth by Fe-deposition and annealing is facilitated on MBE-Bi$_2$Se$_3$ compared to cleaved bulk Bi$_2$Se$_3$ surfaces. FeSe starts to grow already at low temperatures below $T = 100^{\circ}$C with considerably less tetragonal distortion and leaving the surface fully Se-terminated.
We attribute the improved growth conditions to fast diffusion of excess Se adatoms 
towards Bi$_2$Se$_3$ terrace step-edges, where - unlike on atomically flat bulk Bi$_2$Se$_3$ - FeSe islands are preferentially formed. In fact we prove that under these Se-rich conditions Bi remains a spectator element according to the following overall growth equation:
\begin{equation}
\text{Fe} + [\text{Bi}_2 \text{Se}_3]_{n} + [\text{Se}]\rightarrow 
\text{FeSe} + [\text{Bi}_2 \text{Se}_3]_{n}^{*}
\label{eq:reaction}
\end{equation}
The described approach using thin MBE-grown Bi$_2$Se$_3$ substrates of 100~nm scale thickness on insulating BaF$_2$ thus provides the chemically well-defined and continuous FeSe layer system necessary to detect potential interfacial SC properties by lateral transport with a minimum shunt conductivity through the underlying Bi$_2$Se$_3$ topological insulator. 
\begin{figure*}
\includegraphics[width=150mm]{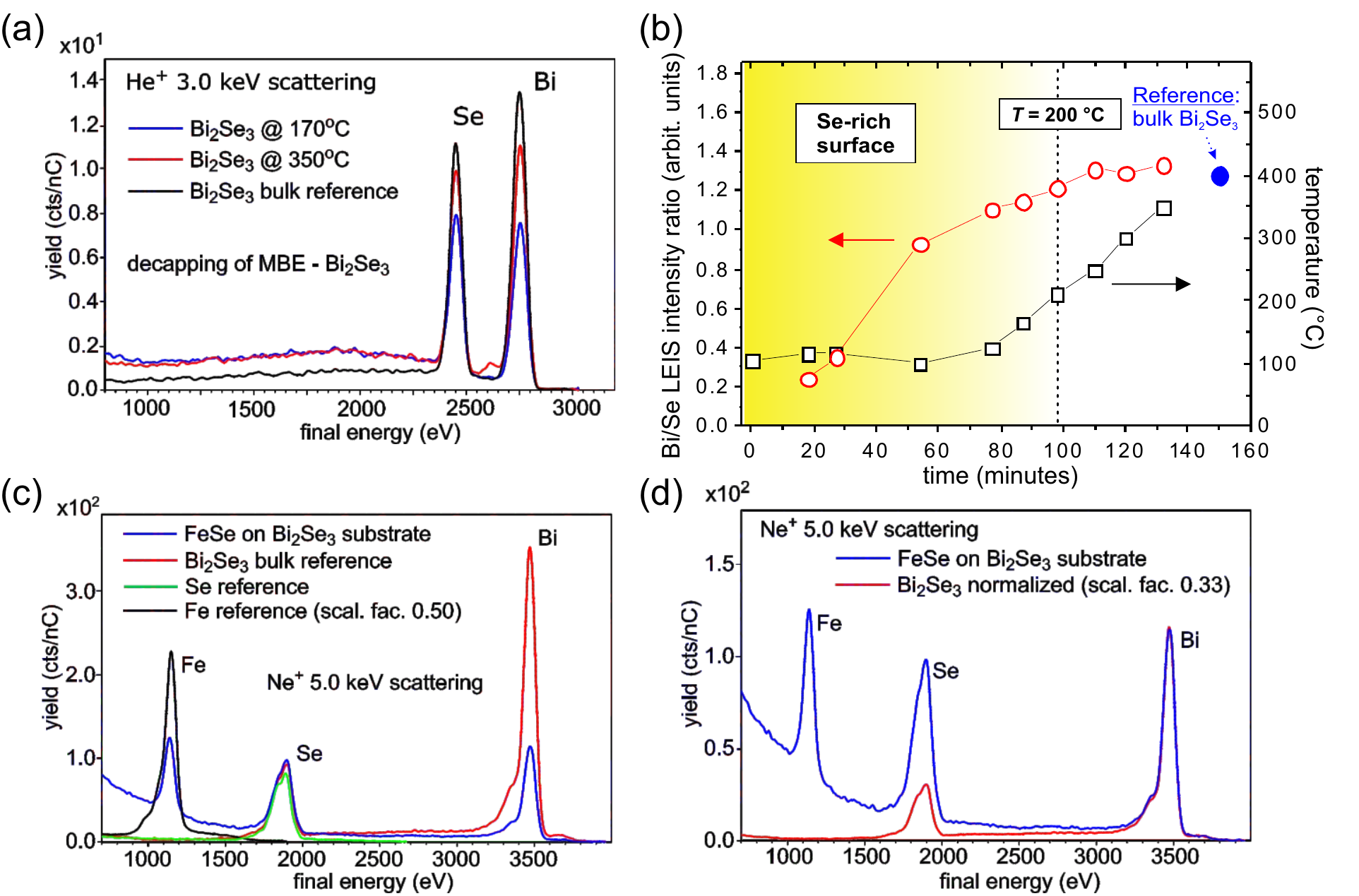} 
\caption{\label{Fig:LEIS} (a) LEIS signals measured during removal of the Se cap. As reference, data from a freshly cleaved bulk Bi$_2$Se$_3$ crystal surface is shown. (b) Temperature dependence of intensity ratios between Bi and Se LEIS signals. (c) and (d) show LEIS specs before and after FeSe growth on decapped Bi$_2$Se$_3$ (respective LEEM and XPS results on the very same sample states are shown in Fig.~\ref{Fig:LEEM} and Fig.~\ref{Fig:XPSCEITEC}. As a reference LEIS spectra of freshly cleaved bulk Bi$_2$Se$_3$, as well as pure Se and Fe films are shown in (c).}
\end{figure*}

\begin{figure*}
\includegraphics[width=110mm]{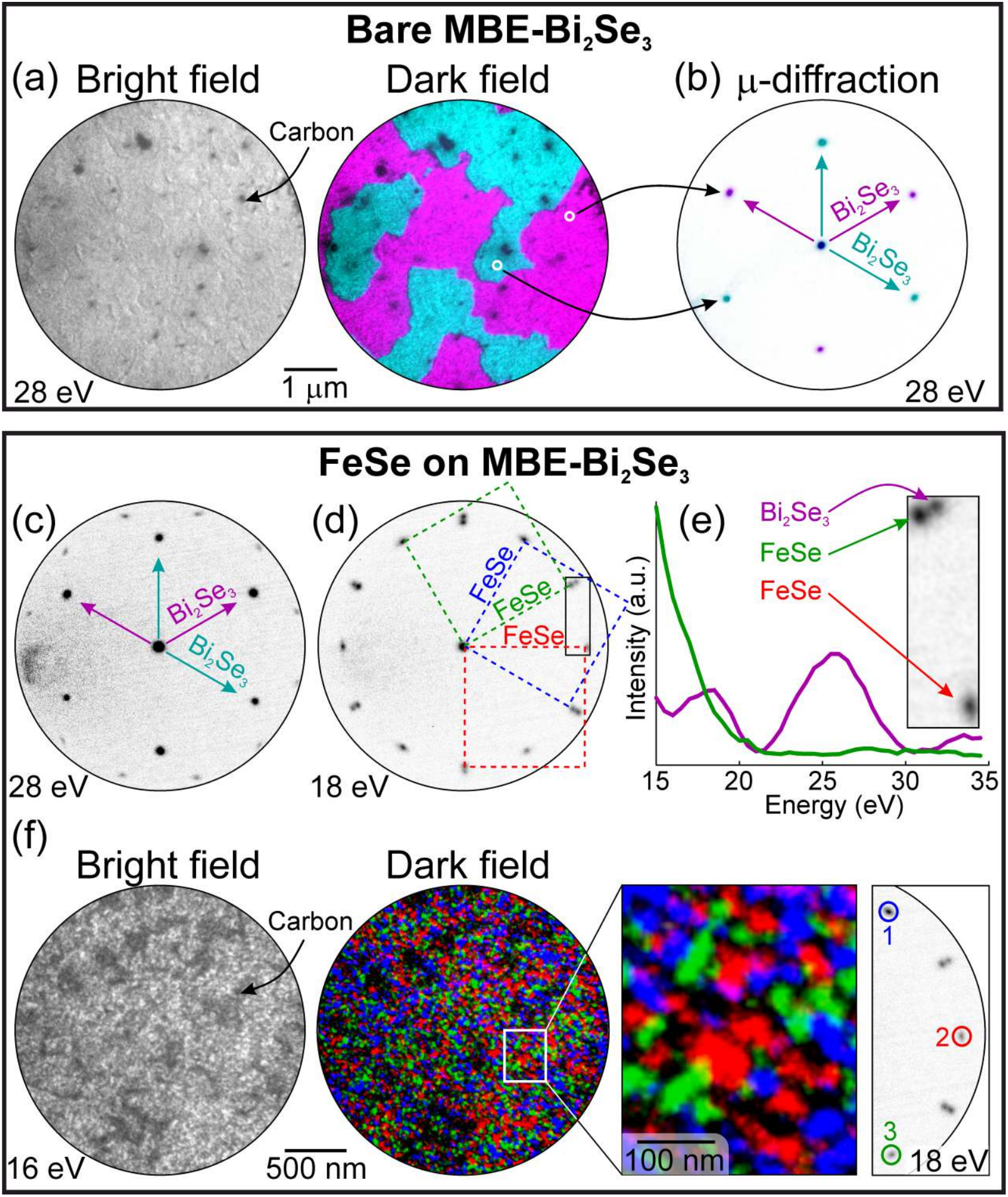} 
\caption{\label{Fig:LEEM} LEEM results in bright- and dark-field real space and  $\mu$-diffraction mode. (a) Bright- and dark-field microscopy of bare Bi$_2$Se$_3$ with (b) respective diffraction signal. (c)-(e) Energy dependent diffraction patterns of FeSe on Bi$_2$Se$_3$ and respective intensities. (f) Bright- and dark-field microscopy of FeSe on Bi$_2$Se$_3$.}
\end{figure*}

\section{Results and Discussion \label{sec:results}}

\begin{figure*}
\includegraphics[width=175mm]{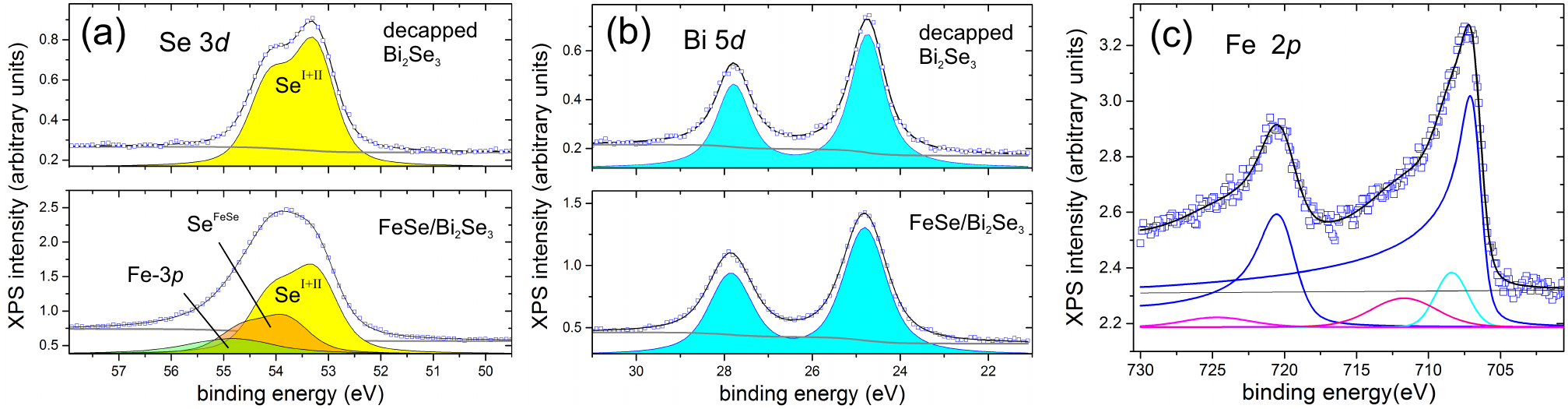} 
\caption{\label{Fig:XPSCEITEC} XPS using Mg K$_{\alpha}$ of bare Bi$_2$Se$_3$ decapped at $T = 160^{\circ}$C ($E_p = 15$~eV) and of FeSe/Bi$_2$Se$_3$ after FeSe growth at $T = 100^{\circ}$C ($E_p = 30$~eV). (a) and (b) shows Se 3$d$ and Bi 5$d$ edges, respectively. (c) shows the Fe 2$p$ edge after FeSe growth. Fits are generated as described in the main text \tcb{and Appendix~\ref{sec:A1}}.}
\end{figure*}

\subsection{Fabrication of continuous FeSe monolayers on micrometer scales: In-situ LEIS, LEEM, and XPS characterisation}
\label{sec:CEITEC}

A thorough understanding of FeSe layer growth procedures ontop of Bi$_2$Se$_3$ requires a careful material characterisation using various analysis techniques. We expect the growth behavior of FeSe monolayers to be highly sensitive to small changes of preparation parameters (temperature calibration, Fe deposition rates, surface contamination,...), which limits the comparability of experimental data achieved for samples prepared in different UHV chambers. To ensure a direct and reliable comparison based on multiple techniques we performed measurements at the material characterisation cluster at {\it CEITEC}, where one and the same sample can be characterised {\it in situ} by RHEED, LEEM, LEIS, and XPS. This Section~\ref{sec:CEITEC} summarizes directly comparable experimental data aquired during different stages of FeSe growth starting from initial decapping of Bi$_2$Se$_3$ surfaces.\\
For a fundamental analysis of growth dynamics of FeSe on Bi$_2$Se$_3$ it is of paramount importance to know the stoichiometry in the topmost surface layer during the different stages of growth. FeSe formation will be governed by atomic diffusion and chemical reaction processes, which depend on the surface properties.\\
LEIS studies have been done to analyze the surface stoichiometries using Helium and Neon ions and results are summarized in Fig.~\ref{Fig:LEIS}. LEIS is the method of choice, since among the various scattering techniques it has the highest possible surface sensitivity and senses mainly the terminating atomic layer of a surface~\cite{Prusa2015}. Depending on the surface atomic geometry and the effectivity of the charge exchange processes between the projectile and surface atoms, the 2$^{\text{nd}}$ atomic layer can also contribute to the surface peak intensity~\cite{BRONGERSMA2007}. Fig.~\ref{Fig:LEIS}(a) and (b) shows LEIS results during the gradual decapping process of Se-capped Bi$_2$Se$_3$ induced by heating. 
The progress of decapping can be quantified by the ratio of integrated Bi and Se signal intensities $r_{\text{Bi/Se}}=I_{\text{Bi}}/I_{\text{Se}}$ in the LEIS spectra, which increases with increasing temperature as shown in Fig.~\ref{Fig:LEIS}(a).
Plotting $r_{\text{Bi/Se}}$ versus time and respective temperatures in Fig.~\ref{Fig:LEIS}(b), at $T = (250 \pm 50)^{\circ}$C values converge to $r_{\text{Bi/Se}}=1.3$, which corresponds well to a reference measurement on {\it in situ} cleaved single bulk Bi$_2$Se$_3$ crystals, which we assume to be Se-terminated. We comment here that Bi-Bi double layer terminated Bi$_2$Se$_3$ was shown not to exhibit any Se LEIS intensity~\cite{He2013-LEIS} and thus can be excluded in our case. This is consistent with our XPS data in Fig.~\ref{Fig:XPSCEITEC}(b) where the Bi 5$d$ peak structure shows a homogeneous shape typical for stoichiometric Bi$_2$Se$_3$. Metallic Bi components e.g. due to Bi-Bi bonding or Bi-Fe coordination have not been detected, which should appear at EB = $24.0$~eV ~\cite{Goncalves2018, Gibson2013, Vondracek2016}.\\
Fig.~\ref{Fig:LEEM}(a) and (b) shows respective LEEM data of the bare Bi$_2$Se$_3$ in the decapped stage. LEEM gives local information on the crystal order on the surface with high spatial resolution. During LEEM experiments the progress of Se decapping procedure was tracked by monitoring the diffraction pattern. The metallic Se is amorphous and thus does not show diffraction. Decapping was stopped when Bi$_2$Se$_3$ diffraction patterns started to appear. It means that according to Fig.~\ref{Fig:LEIS}(b) the decapped Bi$_2$Se$_3$ surface is slightly Se-rich. From XPS data in Fig.~\ref{Fig:XPSCEITEC}(a) we estimate extra Se coverages on the surface to well be below 1~ML, since a metallic Se component expected at 54.8~eV from the cap is not detectable. Instead a Se 3$d$ spectrum typical for clean Bi$_2$Se$_3$ is visible, which can be fitted well by a spin-orbit split doublet Se$^{\text{I+II}}$ plotted in (a), which represents Se in both states Se-I and Se-II in the Se(I)-Bi-Se(II)-Bi-Se(I) sequence of a QL Bi$_2$Se$_3$.\\
Switching between real space Fig.~\ref{Fig:LEEM}(a) and $\mu$-diffraction mode of LEEM at 28~eV in Fig.~\ref{Fig:LEEM}(c) the well-known domain structure in MBE-grown Bi$_2$Se$_3$~\cite{Kriegner2017} can be imaged in the dark field (DF) mode (b) with spatial resolution of 5~nm. Scanning over the domains, the $\mu$-diffraction pattern switches between two 60$^{\circ}$-rotated threefold symmetries of the Bi$_2$Se$_3$(0001) surface. \\
From our results we infer that (i) decapped MBE-grown Bi$_2$Se$_3$ is Se-terminated at decapping temperatures in the range $T = (300 \pm 50)^{\circ}$C and (ii) considerably Se-rich at decapping temperatures below $T = (200 \pm 50)^{\circ}$C as indicated in Fig.~\ref{Fig:LEIS}(b). We expect that Se-rich surface conditions are benefitial to grow high-quality FeSe layers, and help to avoid formation of Bi-containg phases other than that of Bi$_2$Se$_3$.\\ 
The FeSe growth procedure is initiated by Fe deposition at RT (Fe coverage about 1~ML). In LEEM the diffraction patterns immediately disappear due to the disorder introduced by randomely positioned Fe atoms on the surface. During subsequent annealing the formation of ordered FeSe was monitored live in the $\mu$-diffraction mode of LEEM. As shown in Fig.~\ref{Fig:LEEM}(c)-(d) the intensity of the 12-fold pattern is particularily visible at an energy of 18~eV, while for 28~eV the 6-fold pattern of Bi$_2$Se$_3$ is dominant. Rectangular BZs of the three FeSe domains are sketched in (d), and the energy dependence of respective diffraction intensities are compared to that of 6-fold Bi$_2$Se$_3$ in (e). A pronounced oscillatory behavior of the spot intensities due to multiple electron scattering is visible, typical for crystalline systems with high surface order. Looking more closely into the diffraction data at 18~eV, lattice constants of FeSe and Bi$_2$Se$_3$ can be compared. The relative distance of the FeSe reciprocal lattice point with that of Bi$_2$Se$_3$ shown in the inset of (e) is $0.951\pm0.003$. Using the Bi$_2$Se$_3$ in-plane lattice parameter of $a=4.139\,$\AA{} as a reference, we obtain a FeSe lattice parameter of $a=3.769\pm0.012$\,\AA. Within the experimental precision this value corresponds to the bulk FeSe lattice parameter of 3.773\,\AA{}. 
Evaluating the FeSe lattice constant in the direction of the respective 90$^{\circ}$ rotated diffraction spots we find a small average contraction by $(0.5 \pm 0.3)$\%. In comparison to bulk Bi$_2$Se$_3$ (2\% tetragonal distortion) ~\cite{Cavallin2016} FeSe thus grows with significant less distortion on MBE-Bi$_2$Se$_3$. 
From the absence of other diffraction spots we moreover conclude that areal coverages of phases other than Bi$_2$Se$_3$ or FeSe are negligible.\\
LEIS data of the very same FeSe/Bi$_2$Se$_3$ sample state of Fig.~\ref{Fig:LEEM}(c)-(d) are summarized in Fig.~\ref{Fig:LEIS}(c). LEIS shows an additional Fe peak and a reduced Bi intensity with respect to Bi$_2$Se$_3$ as expected for a FeSe layer covering the Bi$_2$Se$_3$ substrate. 
A comparison of LEIS spectra for Neon scattering on FeSe/Bi$_2$Se$_3$, cleaved bulk Bi$_2$Se$_3$, and a pure Se reference sample show equivalent Se peak intensities.
It evidences fully Se-terminated surfaces in all three cases within the resolution of the experiment. This finding is coherent with a Se-terminated Se-Fe-Se trilayer structure expected for tetragonal FeSe, 
Rescaling the LEIS spectra as to match Bi intensities in Fig.~\ref{Fig:LEIS}(d) shows that the Bi surface peak intensity for FeSe/Bi$_2$Se$_3$ is reduced by a factor 0.33 with respect to that of the cleaved bulk Bi$_2$Se$_3$ reference. The shape of the Bi peak remains identical within the experimental resolution. From surface areas terminated by Se-Fe-Se trilayers we expect no Bi signal in LEIS. Thus, 33\% of the surface area gives a Bi signal corresponding to that of Bi$_2$Se$_3$, where Bi occupies the second layer below Se.\\ 
To further clarify the origin the Bi containing phase visible in LEIS, respective XPS spectra in Fig.~\ref{Fig:XPSCEITEC}(a) and (b) give important information of the Bi and Se bonding states. 
Upon FeSe growth we observe strong changes mainly in the Se 3$d$ spectral shape, when compared to Bi$_2$Se$_3$. This is expected, since Se atoms are differently coordinated in FeSe, which will lead to a characteristic chemical shift in XPS compared to Bi$_2$Se$_3$. The main Fe 2$p_{3/2}$ peak position in Fig.~\ref{Fig:XPSCEITEC}(c) is at 707~eV, which corresponds to that of metallic Fe (in covalently bonded FeO or FeS compounds considerably higher values between 709 -712 eV are expected). In Fig.~\ref{Fig:XPSCEITEC}(a) we fit the respctive Se 3$d$ spectrum using an additional Se component Se$^{\text{FeSe}}$ of FeSe and take into account Fe 3$p$ intensity, which is known to appear at EB = 55~eV~\cite{Scholz2012}. \\
In contrast to Se, Bi 5$d$ peaks remains about unchanged and at the same binding energy position, which - as we further prove in Section~\ref{sec:STM} with more surface sensitivity - is due to an intact chemical Bi$_2$Se$_3$ QL structure at the surface and a passive role of Bi in the overall reaction described in Eq.~[\ref{eq:reaction}]. 
The absence of a shoulder in the Bi 5$d$ spectra at EB = $24.0$~eV once more confirms that also after FeSe growth metallic Bi components (e.g. Bi-clusters or from extended Bi-Bi double layers~\cite{Gibson2013}) and metallic Bi-Fe bonding are negligible within the XPS probing depth of about 6~nm~\cite{Valiska2016} (corresponds to 6 QLs of Bi$_2$Se$_3$). Thus, prominent Bi-rich phases like Bi$_4$Se$_3$ (sequence QL-DL-QL-DL ...) and Bi$_1$Se$_1$ (sequence QL-QL-DL-QL-QL-DL ...) can be excluded even if they are covered by a FeSe monolayer.
For a rough estimation of the FeSe thickness we can compare the intensity ratio of the two Se 3$d$ components Se-I and Se-II in Fig.~\ref{Fig:XPSCEITEC}(a). Following Seah et al.~\cite{Seah2002}, we estimate the average FeSe thickness to be $(7.5\pm 1.0)$~\AA, which corresponds to 1.3 UCs. 
The areal distribution of FeSe and Bi$_2$Se$_3$ coverages can identified by dark field (DF) imaging shown in Fig.~\ref{Fig:LEEM}(f). For DF imaging, refractive spots of the three FeSe domains [see (f), right panel] were chosen. The DF image reveals a carpet of three FeSe domain orientations with approximately equal areal coverage. Typical domain sizes are 50~nm and shapes appear elongated along the high-symmetry directions and step-edges of the underlying Bi$_2$Se$_3$, an effect which will be discussed in detail on the basis of STM data in Section~\ref{sec:STM}. From the spatial resolution 5~nm in the DF mode we estimate a total areal FeSe coverage of $(72\pm 5)$\% in excellent agreement with the LEIS estimation of $(67\pm 3)$\% . 

From the versatile combination of LEEM, XPS, and LEIS studies described in this Section~\ref{sec:CEITEC} we confirm the high crystalline order in our FeSe layers with large homogeneous coverage of mostly coalescing domains. We attribute the absence of pronounced Bi-rich Bi$_n$Se$_m$ phases to the Se-rich surface conditions during FeSe growth, which suppresses the formation of other phases as discussed in Section~\ref{sec:STM}. 


\subsection{Electronic band structure characterisation of the FeSe-Bi$_2$Se$_3$ interface}
\label{sec:ARPES}
\begin{figure*}
\includegraphics[width=115mm]{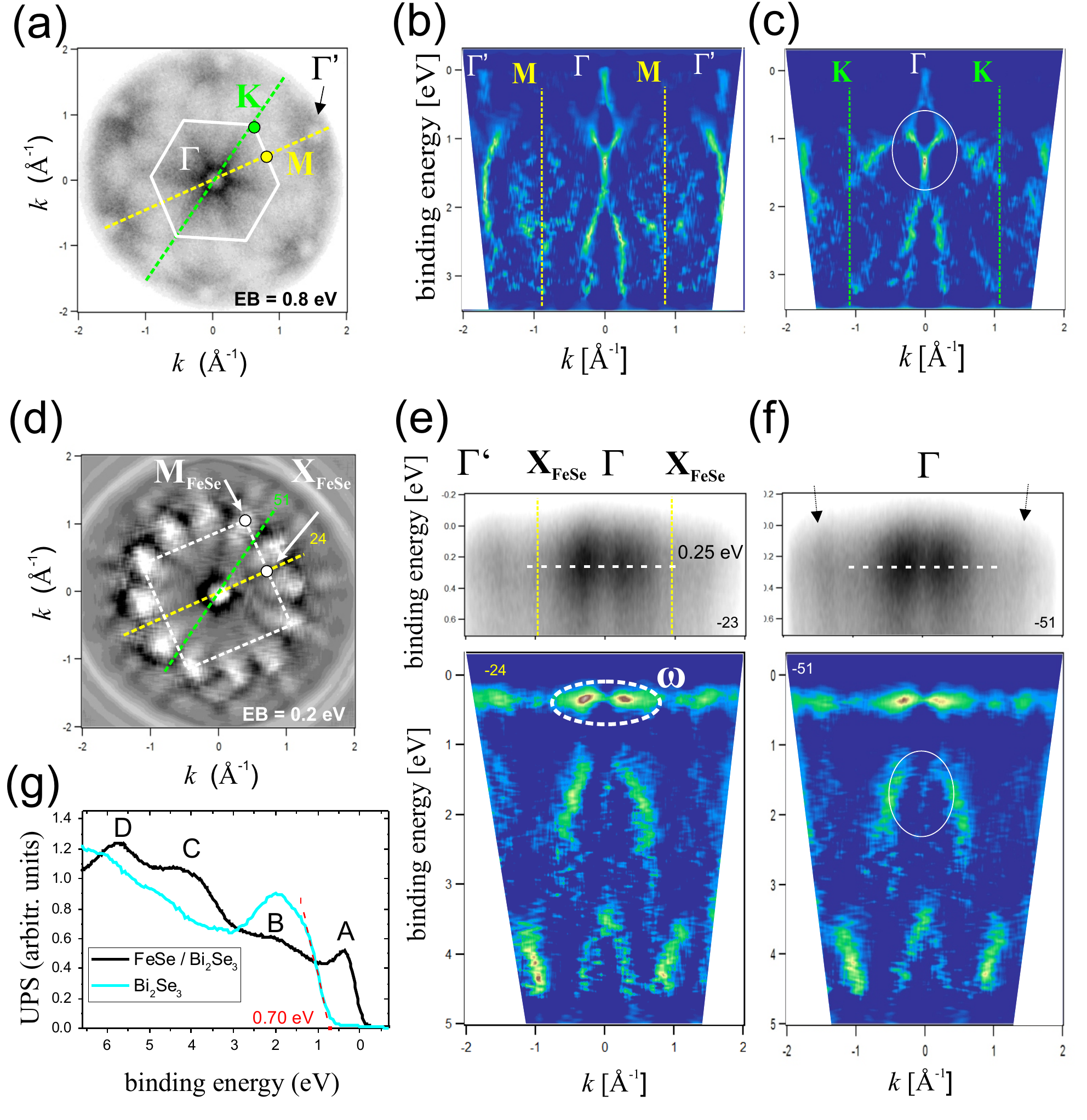} 
\caption{\label{Fig:kPEEM} $k$PEEM data at $h\nu = 21.2$eV of (a)-(c) bare Bi$_2$Se$_3$ decapped at $T = 180^{\circ}$C and (d) - (f) FeSe/Bi$_2$Se$_3$ after FeSe growth at $T = 320^{\circ}$C. (g) Plots of the $k$-space integrated UPS intensities of Bi$_2$Se$_3$ and FeSe/Bi$_2$Se$_3$.}
\end{figure*}

\begin{figure*}
\includegraphics[width=175mm]{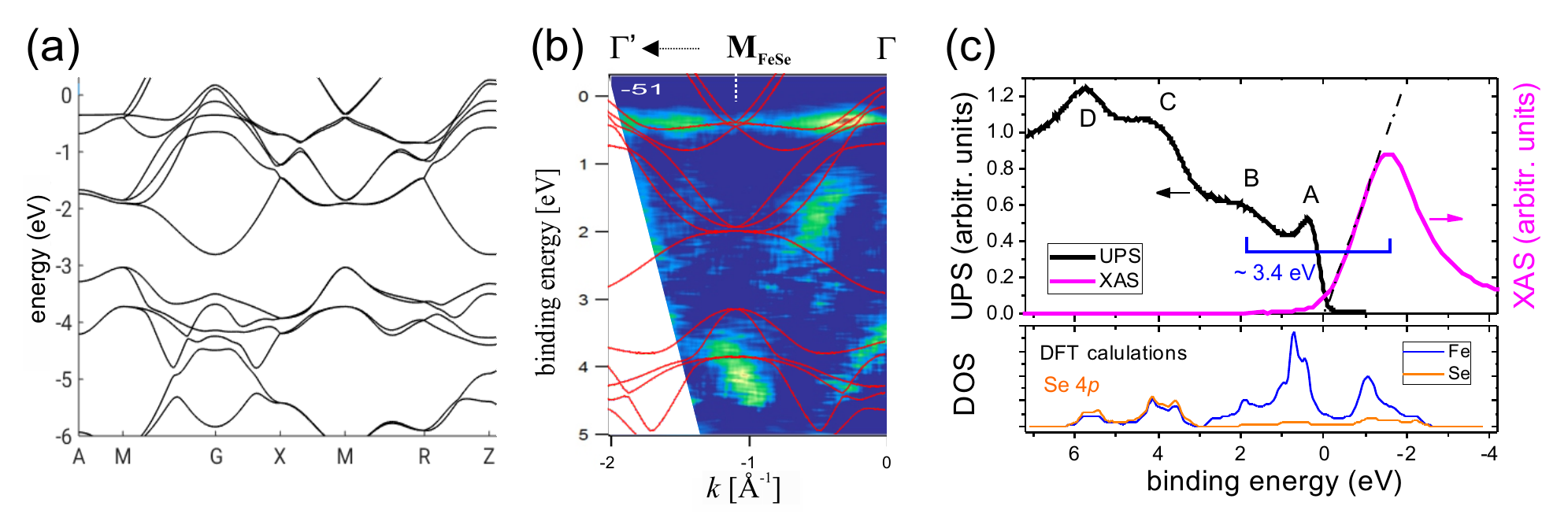} 
\caption{\label{Fig:kPEEM+theory} (a) DFT LSDA + LSDA calculation of the bulk FeSe band structure. (b) Experimental $k$-PEEM data at $h\nu = 21.2$eV of Fig.\ref{Fig:kPEEM}(e) with DFT along $\Gamma$-M. (c) $k$-space integrated UPS of FeSe/Bi$_2$Se$_3$ and Fe $L_{3,2}$ XAS data. The energy alignment of the XAS data was done by matching the $L_{3}$ leading edge with EB = 0 (see dash-dotted line). In the lower panel projections of the total DOS on Fe and Se states are shown for comparison.}
\end{figure*}

\begin{figure}
\includegraphics[width=85mm]{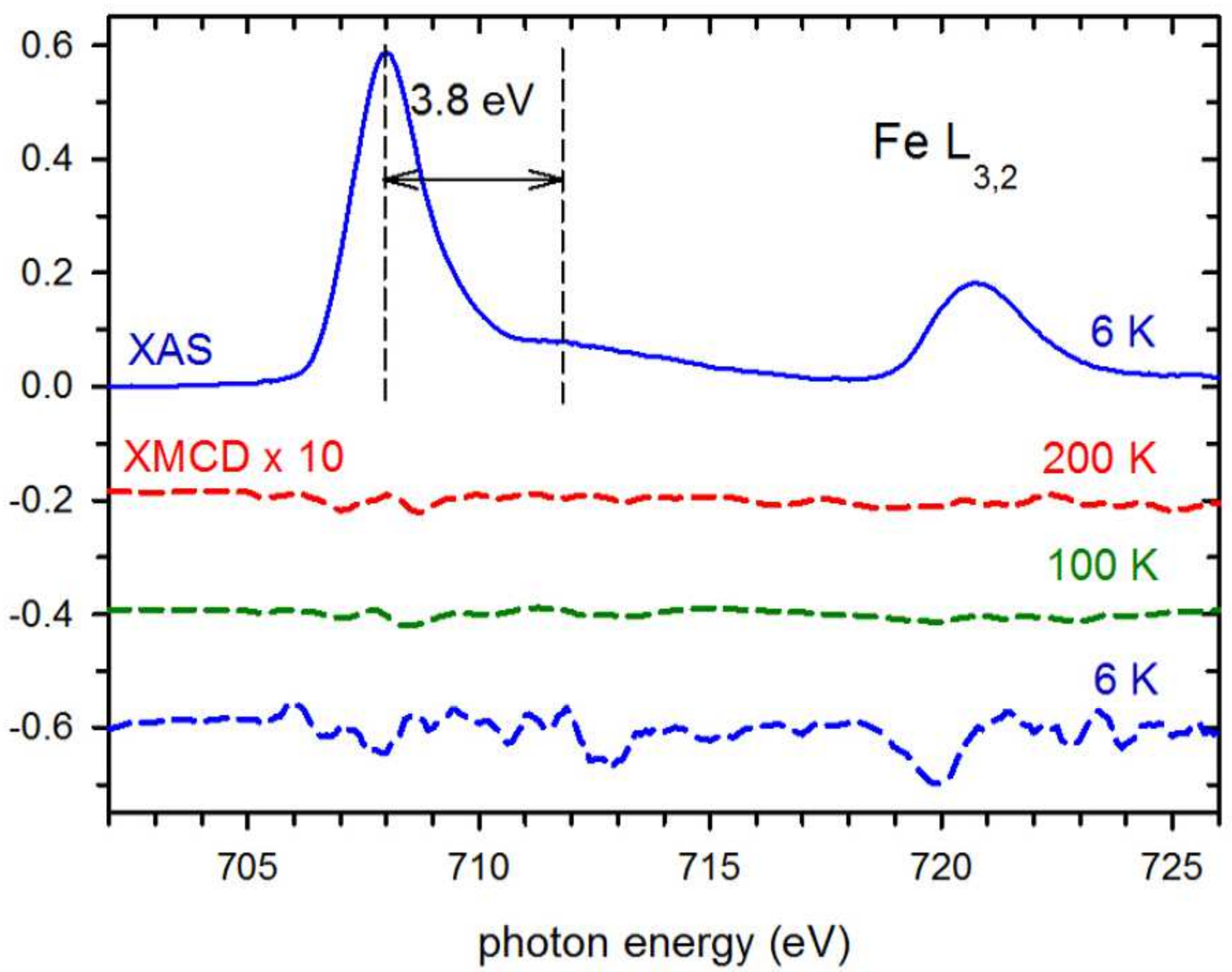} 
\caption{\label{Fig:XMCD} XAS and recpetive XMCD spectra of FeSe on Bi$_2$Se$_3$ taken at temperatures $T=200, 100$ and $6$K in polar magnetic fields $B=0.2$~T and $\Theta = 0^{\circ}$. XMCD signals are enlarged by a factor 10. }
\end{figure}

Figure~\ref{Fig:kPEEM}(a) displays a $(k_x, k_y)$ ARPES distribution of decapped Bi$_2$Se$_3$ at EB $=0.8$~eV, which corresponds to an energy cut just below the VBM maximum located at EB $=0.7$~eV as derived from UPS data in Figure~\ref{Fig:kPEEM}(g)]. The $k$-space range covers the full trigonal 1$^{\text{st}}$ BZ around $\bar{\Gamma}$ (marked in blue) and extends beyond the six $\bar{\Gamma}$ points from the 2$^{\text{nd}}$ BZs (denoted $\bar{\Gamma}'$). A 6-fold symmetry of ARPES patterns is expected from the UV light spot size of several hundreds of micrometers, which averages over many equally prominent $60^{\circ}$ Bi$_2$Se$_3$ domains observed by DF LEEM described in Section~\ref{sec:CEITEC}. $(k,E)$-cuts in the high symmetry directions $\bar{\Gamma}-\bar{M}$ and $\bar{\Gamma}-\bar{K}$ are plotted in (b) and (c) with 2$^{\text{nd}}$ derivative contrast. 
At $\bar{\Gamma}$ and $\bar{\Gamma}'$ the $n$-doped topological surface states (TSSs) are visible with a Dirac point (DP) located $0.2$~eV below $E_F$. They appear as a weak plateau-like intensity in the integrated UPS signal above the Bi$_2$Se$_3$ VBM [see Fig.~\ref{Fig:kPEEM}(g)]. We comment here that ARPES data on decapped Bi$_2$Se$_3$ samples is generally found to very homogeneous over the whole sample surface.\\ 
Figure~\ref{Fig:kPEEM}(d)-(f) summarizes respective photoemission data after FeSe growth (Fe deposition of approximately 1~ML coverage at room temperature and subsequent annealing at $T = 320^{\circ}$C for 30~min). By XPS we verified {\it in situ} that this sample has the same FeSe coverage as the one characterized in Section~\ref{sec:CEITEC}. A $(k_x, k_y)$-cut at EB $=0.2$~eV shows the characteristic 12-fold symmetric $k$-space pattern of the three equally prominent FeSe domains [Fig.~\ref{Fig:kPEEM}(d)]. The BZ of one domain is marked as a square with according high symmetry directions $\bar{M}_{\text{FeSe}}$ and $\bar{X}_{\text{FeSe}}$. $(k,E)$ cuts are shown in Fig.~\ref{Fig:kPEEM}(e) and (f) [same directions as those shown in (b) and (c), respectively], where the former corresponds to a cut along $\bar{\Gamma}-\bar{X}_{\text{FeSe}}$ of the indicated FeSe domain BZ in (d). 
We comment that of course these $(k,E)$-cuts contain intensities also from the other two FeSe domains, which leads to a
mixing of three $(k,E)$-cuts along directions $\bar{\Gamma}-\bar{X}_{\text{FeSe}}$ and $\pm 12^{\circ}$ rotated $\bar{\Gamma}-\bar{M}_{\text{FeSe}}$ as discussed e.g. in Eich et al. for FeSe partially covering bulk Bi$_2$Se$_3$ ~\cite{Eich2016}. In contrast to Eich et al. however we do not observe traces of the Bi$_2$Se$_3$ band structure as expected for the continous coverage of Bi$_2$Se$_3$ by FeSe in our case [see Section~\ref{sec:CEITEC}].  \\ 
Our photoemission spectroscopy data of a monolayer FeSe on MBE-grown Bi$_2$Se$_3$ exhibit several typical band structure features of FeSe. In contrast to decapped Bi$_2$Se$_3$ the UPS shown in Fig.~\ref{Fig:kPEEM}(g) has a metal-like large spectral weight up to the Fermi level and shows
four characteristic peaks indicated as A, B, C, and D in the range up to 6~eV. These peaks are also observed in bulk FeSe photoemission spectrsocopy~\cite{Liu_2017, Takayoshi2012} and can be assigned to Fe 3$d$ (A and B) and Se $4p$ (C and D) dominated states as we will discuss further below.
Also $(k,E)$-cuts clearly reveal several characteritic features known from the FeSe band structure, such as the flat $\omega$ band at EB $=0.25$~eV around $\bar{\Gamma}$ [see also theory Ref.~\cite{Bazhirov_2013, Zheng2013}] and strongly dispersing bands in the energy range EB $=1.5-2.5$~eV and EB $=3.5-5.0$~eV. The former dispersing bands appear broad supporting the intepretation as Hubbard-like bands due to incoherent many-body excitations in Fe 3$d$ states~\cite{Watson2017}.\\
For a quantitative interpretation we performed DFT band structure calculations shown in Fig.~\ref{Fig:kPEEM+theory}(a). Despite our LMTO-DFT approach assuming a bulk FeSe crystal structure the calculations reproduce the experiment rather well at larger EBs above $=1$~eV as shown in Fig.~\ref{Fig:kPEEM+theory}(b) for the $\bar{\Gamma}-\bar{M}_{\text{FeSe}}$ direction, where the states are of Se 4$p$ character. The good agreement is visible also by comparing the UPS data with the calculated DOS in Fig.~\ref{Fig:kPEEM+theory}(c), where peaks B, C and D are reproduced rather well. The energy position of the sharp peak A however appears closer to $E_F$ in the experiment compared to theory. Indeed, DMFT methods were reported to be more suitable to reproduce low-energy band structure as they seem capture local Coulomb interactions and Hund`s coupling effects Ref.~\cite{Watson2017}.\\
Our calculations predict electron and hole pockets at $\bar{M}$ and $\bar{\Gamma}$-points [see Fig.~\ref{Fig:kPEEM+theory} (a) and (b)]. In the upper panels of (e) and (f) the band structure close to the Fermi level is measured with better statistics. Close to $\bar{\Gamma}'$ we observe strongly dispersing bands at $k$-values around $\pm 1.3$ as indicated by arrows in Fig.~\ref{Fig:kPEEM}(f), which we asign to bands forming the hole pockets at $\bar{\Gamma}'$ according Fig.~\ref{Fig:kPEEM+theory}(a). They are weakly visible also at $\bar{\Gamma}$ but are overshadowed by the broad strong intensity of the $\omega$ band in the 1st BZ. Respective electron pockets predicted at the $\bar{M}$ points are not visible in the respective data but may be visible in 2nd derivative contrast in Fig.~\ref{Fig:kPEEM+theory}(b) at $\pm 1.2$ where the $\bar{M}$ is expected.\\
The band structure calculations in Fig.~\ref{Fig:kPEEM+theory}(c) predict a large DOS at energies 1-2~eV above the Fermi level, which is largely composed of Fe 3$d$ states . To probe these unoccupied states close to the Fermi level we performed x-ray absorption spectroscopy (XAS) measurements at the Fe $L_{3,2}$ edges,
which correspond to dipole transitions from localized spin-orbit split 2$p$ core levels (2$p_{1/2}$ and 2$p_{3/2}$) into empty 3$d$ states (2$p^6$3$d^n \rightarrow 2p^5$3$d^{n+1}$). We stress that XAS data provides information on empty Fe states in the presence of a 2$p$ core hole. Its spectral shape does not reflect the empty DOS in a direct way but depends to orbital selective dipole selection rules and reflects collective local screening effects and Fe valance and spin states.  \\  
In Fig.~\ref{Fig:XMCD} we summarize temperature dependent XAS studies of FeSe/Bi$_2$Se$_3$ at the Fe $L_{3,2}$ edges left (+) and right (-) circularly polarized light in polar geometry ($\Theta = 0^{\circ}$). Respective temperature dependent XMCD spectra in small fields of 0.2~T show no sign of magnetic response as expected for a non-ferromagnetic system. The non-dichroic XAS spectra (XAS$^{+}$ + XAS$^{-}$)  exhibit no sharp multiplet structures, and - similar to bulk Fe - the $L_{3}$ peak is located at 708~eV. Core-hole interactions in the $2p^53d^{n+1}$ state are thus considerably screened by free carriers. However, compared to metallic Fe, $L_{3}$ and $L_{2}$ resonances are narrower 
and with a prominent shoulder at about 712~eV, which is 3.8~eV from the main peak as indicated in Fig.~\ref{Fig:XMCD}. It should be noted that a similar shoulder separated by 4.5~eV was observed also by direct probing of the unoccupied DOS of bulk FeSe by light inverse photoemission spectroscopy in the UV range~\cite{Yokoya2012}. 
For bulk Fe-chalcogenides like FeSe~\cite{Saini2011} or CsFe$_2$Se$3$~\cite{Takubo2017} this shoulder in XAS can be understood within a FeCh$_4$ cluster model taking into account 3$d$-3$d$ Coulomb energy ($U=1.5$~eV) and charge transfer between 4$p$-ligands and 3$d$ orbitals (Slater-Koster parameters $pd\sigma = -1.5$ and $pd\pi=+0.69$~eV for hybridisation terms between $d^6$ and 3$d^7 L$ configurations)~\cite{Saini2011}. We extract XAS branching ratios $L_{3}/L_{3,2}$ of $(0.72\pm 0.01)$ [$L_{3}/L_{2}$ of $(2.57\pm 0.03)$] suggesting a typical Hunds rule high-spin state~\cite{Thole88, Honolka2009}. The position of the Fermi level in our calculated Fe DOS in Fig.~\ref{Fig:kPEEM+theory}(c) predicts an occupancy of 6.7 which is in line with Fe configurations between $d^6$ and 3$d^7$. 
It is worth to mention that within this model the missing $L_{3}$ shoulder in the non-SC bulk FeTe counterpart can be interpreted as a reduced 3$d$-4$p$ hybridisation due to longer Fe-Ch bond in the case of Te.\\
Despite the larger complexity of XAS compared to photoemsission data, we can assume that in the case of metallic FeSe with well-screened core holes XAS spectra reflect the final state DOS rather well. In Fig.~\ref{Fig:kPEEM+theory}(c) we therefore added $L_{3}$ XAS data at negative EB values. Since the raising edge of $L_{3}$ cannot be used to define the $E_F$ position in a straight forward manner (it cannot be fitted by a temperature dependent Fermi function) we use a tangential leading edge method to adjust the onset of XAS with $E_F = 0$ defined by UPS as shown in Fig.~\ref{Fig:kPEEM+theory}(c). Using this method we find a surprisingly good agreement of the XAS spectral shape and energy position with inverse photoemission data of bulk FeSe~\cite{Yokoya2012} and also our DOS calculations. The energy difference between the Fe $L_{3}$ peak with the characteristic feature B in UPS amounts to 3.4~eV, which is close to the value 3.5~eV found in the comparison of photoemission and inverse photoemission data where 3.5~eV was reported~\cite{Yokoya2012}. It suggests that in the well-screened case of FeSe the XAS spectral shape does give a reasonable estimation of the unoccupied Fe 3$d$ dominated DOS.\\
\subsection{Step-edge assisted FeSe island growth dynamics}
\label{sec:STM}

\begin{figure*}
\includegraphics[width=175mm]{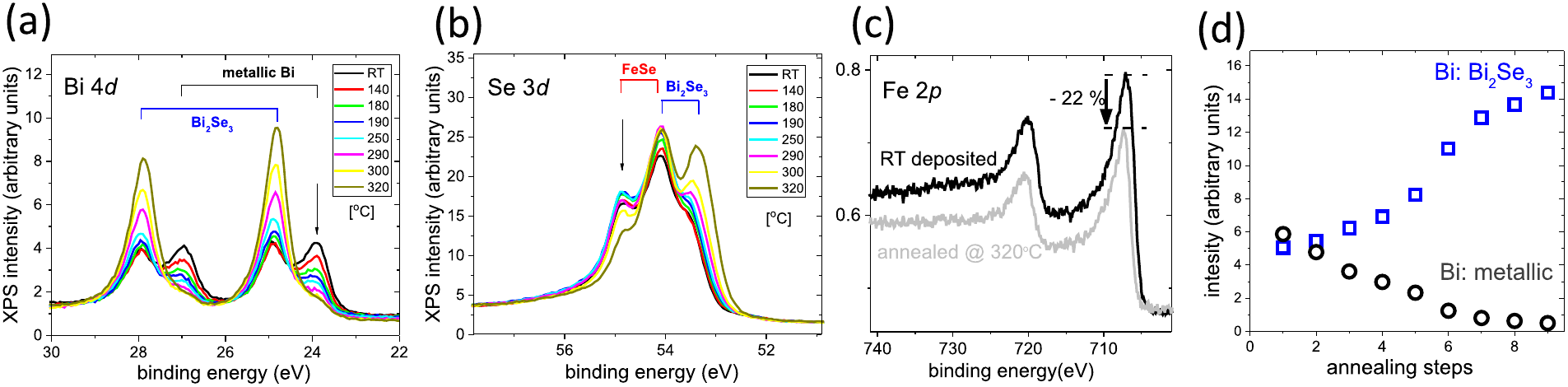} 
\caption{\label{Fig:UPS} Monitoring FeSe formation in UPS at maximum surface sensitivity. Measurements are done at photon energies $h\nu$ = 130~eV during an annealing cycle up to $T = (320 \pm 50)^{\circ}$C. (a) and (b) show Bi 5$d$ and Se 3$d$ CL spectra during progressive annealing. (c) Fe 2$p$ intensity before and after annealing at the maximum temperature $T = (320 \pm 50)^{\circ}$C. (d) Development of the temperature dependent spectral weight of Bi$_2$Se$_3$, FeSe, and metallic Bi components.}
\end{figure*}

\begin{figure}
\includegraphics[width=85mm]{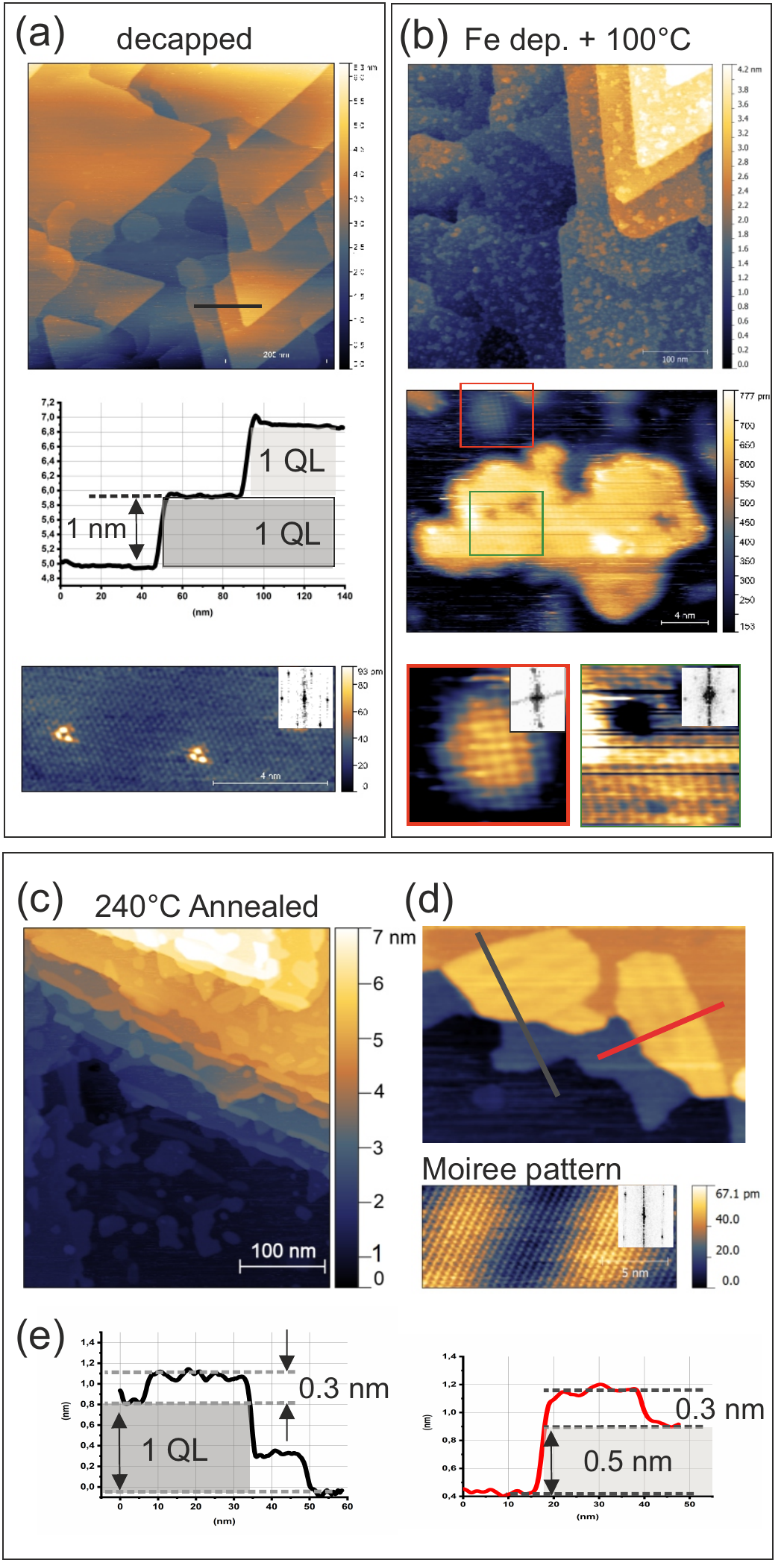} 
\caption{\label{Fig:STM} STM images at different stages of FeSe preparation. Insets show respective LEED images and FFT-analysis. (a) Bi$_2$Se$_3$ decapped at $T = (190 \pm 30)^{\circ}$C. (b) After 0.7 ML Fe deposition and annealing at $(100 \pm 30)^{\circ}$C. (c) and (d) After annealing at $(240 \pm 50)^{\circ}$C.}
\end{figure}

From the similarity of Bi CL spectra before and after FeSe growth on Bi$_2$Se$_3$ (see Fig.~\ref{Fig:XPSCEITEC}) we concluded that Bi - in total - plays the role of a spectator element in the FeSe growth process. This finding is surprising since Fe atoms deposited on Bi$_2$Se$_3$ and Bi$_2$Te$_3$ surfaces both at room temperature or low temperatures are known to induce characteristic shoulders at EBs of about 24~eV in the Bi 5$d$ edges~\cite{Vondracek2016, Scholz2012}, evidencing considerable interaction between Fe and Bi. In the following we will show that such shoulders are visible in a transient state during the FeSe growth process but disappear when FeSe has fully developed. Our findings give important insight into the FeSe growth dynamics at the atomic scale.\\
In Fig.~\ref{Fig:UPS} (a) and (b) Bi 5$d$ and Se 3$d$ edges are shown after 1~ML Fe deposition at room temperature and during subsequent annealing up to temperatures $T = (320 \pm 50)^{\circ}$C. CLs are measured at ELETTRA with highest surface sensitivity at photon energies $h\nu$ = 130~eV (probing depth of about 5~\AA). Before annealing the Bi 5$d$ CL shape confirms the expected charateristic shoulder at 24~eV corresponding to a metallic-like Bi state induced by the presence of Fe, which is a consequence of the known heavy relaxation of 3$d$ surface atoms into the Bi$_2$Se$_3$ surface~\cite{Vondracek2016}. Due to the high surface sensitivity only half of the Bi spectral intensity remains at a EB corresponding to that of the Bi$_2$Se$_3$ host state. Surprisingly with the formation of FeSe during annealing the metallic state of Bi disappears entirely, while Fe 2$p$ CLs intensities remain comparable [see Fig.~\ref{Fig:UPS}(c)]. Fitting the Bi 5$d$ by two components [see Section~\ref{sec:CEITEC}] the transfer of spectral weight from the metallic state to that of Bi$_2$Se$_3$ can be quantified in Fig.~\ref{Fig:UPS}(d). The original Bi chemical state corresponding to Bi$_2$Se$_3$ is thus fully restored when the FeSe growth is completed at about $T = (320 \pm 50)^{\circ}$C, which implies an intact QL structure in the final state (removing one Se layer from a QL would produce a chemical shift in the adjacent Bi layer) according to Eq.~\ref{eq:reaction}. This is in line with our more bulk sensitive XPS measurements in Fig.~\ref{Fig:XPSCEITEC} with a probing depth of about 15-20~\AA.\\
Looking more closely at the Se 3$d$ spectrum we see that after Fe deposition the spectrum is complex but can be fitted by two main components representing the Se$^{\text{I+II}}$ components of Bi$_2$Se$_3$ and an additional Se$^{\text{FeSe}}$, in line with Fig.~\ref{Fig:XPSCEITEC}(a) and respective discussion. 
Surprisingly, our analysis suggests that the Se$^{\text{FeSe}}$ component is present down to temperatures well below 100~K.\\
In order to verify the formation of FeSe seeds already at low temperatures we performed STM measurements at various stages of FeSe growth. Figure~\ref{Fig:STM} summarizes the studies. Fig.~\ref{Fig:STM}(a) shows the typical morphology of a decapped bare MBE-grown Bi$_2$Se$_3$ surface, which consists of triangular shaped pyramids and a high density of QL steps of 1~nm height.\\ 
After Fe deposition at room temperature, STM shows rough profiles without atomic resolution, which we attribute to the random position of Fe on the surface and the reported partial diffusion into the first QL~\cite{Schlenk2013}. Already after annealing at $T = 100 ^{\circ}$C, however, small FeSe seeds of approximately 5~nm lateral size become visible [see in Fig.~\ref{Fig:STM}(b)]. The seeds show the characteristic $3.7$\AA{} cubic atomic lattice with a typical height of $(1.6\pm 0.2)$\AA{} with respect to the Bi$_2$Se$_3$ terraces but without a Moir\'{e}e pattern. 
We thus have confirmed our UPS results showing the presence of Se$^{\text{FeSe}}$ at comparably low temperatures.\\
The surface structure after annealing at $T = (240 \pm 50)^{\circ}$C has changed in several ways. Large FeSe islands have formed [see Fig.~\ref{Fig:STM}(c)-(e)] (typical size of 50~nm similar to those overved in dark field LEEM in Fig.~\ref{Fig:LEEM}), which exhibit the characteristic Moir\'{e}e pattern reported first in 2016 by Cavallin et al~\cite{Cavallin2016} on bulk Bi$_2$Se$_3$. In our case of MBE-grown Bi$_2$Se$_3$ surfaces, 
they grow preferentially along the QL step-edges in an elongated fashion [see Fig.~\ref{Fig:STM}(c)] both on the higher and lower part of the QL step as shown by the line scans in Fig.~\ref{Fig:STM}(e). With a typical height of $(3.0\pm 0.2)$\AA{} with respect to Bi$_2$Se$_3$ terraces these islands appear much higher than the low-temperature FeSe seeds. The Moir\'{e}e height variation $\pm 0.25$\AA{} shown in Figure~\ref{Fig:STM}(e) coincides well with previoulsy reported ones on bulk Bi$_2$Se$_3$~\cite{Cavallin2016, Eich2016}.\\
In the following we discuss the origin of the observed preferential formation of FeSe along Bi$_2$Se$_3$ step-edges. Island growth kinetics on surfaces is known to be determined by parameters of atom interlayer and terrace diffusion and nucleation probabilities. Atomic diffusion processes at and across Bi$_2$Se$_3$ QL step-edges are especially complex, as the steps are composed of several Se and Bi atomic layers and thus cannot be modeled as a simple 1D diffusion barrier. Recent calculations e.g. propose that the efficient layer-by-layer growth of Bi$_2$Se$_3$ during MBE may be due to a step-edge heteroatomic barrier reduction mechanism, which e.g. reduce diffusion barriers for Se adatoms across QL barriers~\cite{Kim2019}. As a possible scenario for the step-assisted FeSe growth we propose that randomly room-temperature deposited Fe atoms start to migrate at elevated temperatures and get preferentially pinned at QL steps at the bottom (by high step-edge detachment barriers) but also the top (limited by a high Schwoebel barrier). Schwoebel barriers can be considered high since in STM the FeSe coverage seems to scale with the terrace signifying little Fe diffusion across step edges. During Fe migration Se-rich surfaces conditions of MBE-Bi$_2$Se$_3$ discussiond in Section~\ref{Fig:LEIS} and respective fast Se adatom surface diffusion provide the necessary local excess Se concentration at the step-edges to form FeSe seeds while keeping the suporting Bi$_2$Se$_3$ terrace intact.

\section{Conclusions \label{sec:conclusions}}

We present a detailed study of FeSe growth on MBE-grown Bi$_2$Se$_3$ substrates initiated by Fe deposition (about 1 monolayer coverage) and subsequent annealing at elevated temperatures between $100^{\circ}$C and $320^{\circ}$C. The FeSe growth method was first reported by Cavallin et al.~\cite{Cavallin2016} in 2016 on bulk Bi$_2$Se$_3$ surfaces, however FeSe coverage and the surface homogeneity after growth has been under debate especially regarding the possible occurrance of Bi-rich phases. Using various surface sensitive techniques we show that - in contrast to bulk Bi$_2$Se$_3$ surfaces - MBE-grown Bi$_2$Se$_3$ thin film substrates with a high density of 1-nm high step-edges are perfect seeding grounds for FeSe monolayers with high coverage and high chemical order. The resulting FeSe domains of typically 30-50~nm lateral size exceeding the typical Cooper pair length scale and form a coalescing carpet ontop of Bi$_2$Se$_3$. This makes these layers a perfect system to study superconducting properties in lateral transport experiments.\\ 
A thorough characterisation by photoemission and x-ray absorption spectroscopy at the Fe $L_{3,2}$ edges is presented, which confirms the characteristic features of the FeSe band structure and a high-spin state corresponding to an Fe occupancy between $d^6$ and $d^7$. Results are in line with here presented density functional theory band structure calculations. \\
We attribute the facilitated growth on MBE-grown Bi$_2$Se$_3$ to preferential nucleation of Fe atoms along the many step-edges which are absent in respective cleaved atomically flat bulk Bi$_2$Se$_3$ substrates. In the case of MBE-grown Bi$_2$Se$_3$ surfaces we detect high Se excess adatom concentrations which are necessary for the formation of well-odered FeSe islands ontop of Bi$_2$Se$_3$. Under these conditions the underlying Bi$_2$Se$_3$ terrace stoichiometry is not compromised during growth, leaving the surface Se-terminated at all stages of growth and without the formation of Bi-rich phases.   \\

\section{Methods \label{sec:methods}}
Bi$_{2}$Se$_{3}$(0001) films were epitaxially grown on insulating
BaF$_2$(111) and the structural order of the material was monitored by reflection high energy electron diffraction as described in earlier works~\cite{Valiska2016}. The epitaxial heterostructures were finally protected
against oxidation by amorphous metallic Se cap layers, which allows to recover clean surfaces by an {\it in-situ} decapping procedure under ultra-high vacuum (UHV) conditions. FeSe growth on decapped Bi$_{2}$Se$_{3}$(0001) surfaces were done using a two-step procedure: (i) deposition of Fe coverages in the range 0.3 - 1.0 monolayers (ML) at room temperature (RT) and (ii) subsequent annealing at elevated temperatures. Fe evaporator fluxes were calibrated using a quartz balcance and by comparison of Fe and (Bi, Se) XPS intensities, which were cross-referenced to coverages estimated from STM data.
 
At the at {\it CEITEC} Nano Core Facility one and the same sample can be characterised {\it in situ} by various surface sensitive techniqes including low-energy electron microscopy (LEEM), low-energy ion scattering (LEIS), and x-ray photoelectron spectroscopy (XPS).
LEEM/$\mu$-LEED, LEIS and XPS measurements were performed in a complex ultrahigh vacuum (UHV) system with a FE-LEEM P90 (Specs GmbH), Qtac 100 (ION-TOF GmbH), and Phoibos 150 spectrometer, respectively. Here, the samples were prepared in a separate UHV chamber and moved to the instruments for analysis via a linear transfer system under UHV conditions; the base pressure in all chambers was below $2\times 10^{−10}$~mbar. Bright field LEEM was done with electrons in the energy range [0, 35]~eV from the (0,0) diffracted beam. The diffraction and $\mu$-diffraction patterns were collected from spots with elliptical shape of $15\times 10$~$\mu$m$^2$ and circular area with diameter of 185~nm. For LEIS measurements the normal incidence primary beam of Helium or Neon ions with characteristic diameter 50~$\mu$m was scanned over the surface areas of about $2\times 2$~mm$^2$. The double-thoroidal energy analyzer with full azimuthal acceptance maximizes the sensitivity of the instrument. This enables to minimize the surface modification during the measurement (static measurement conditions). Typical ion fluences for a single spectrum was $1.2\times 10^{14}$ He$^{+}$ ions/cm$^2$ and $0.3\times 10^{14}$ Ne$^{+}$ ions/cm$^2$. The scattering angle is fixed to 145$^{\circ}$. XPS experiments at {\it CEITEC} were performed by non-monochromatized Mg K$_{\alpha}$ radiation and normal emission geometry (emission angle 0$^{\circ}$) at pass energies $E_p$ of 15~eV and 30~eV.

$k$-PEEM measurements with laboratory light sources were done using an {\it Omicron NanoESCA}
instrument located at the SAFMAT center in Prague. The photoemission spectrometer is based on a PEEM column and a imaging double hemispherical energy filter. A transfer lens in the electron optics switches between real space and $k$-PEEM mode, which also allows to do classical XPS with monochromatized
Al K$_{\alpha}$ radiation. Energy dependent $k$-space mapping of the Brillouin zone (BZ) was performed using a helium discharge lamp at h$\nu$ = 21.2 eV with an analyser kinetic energy resolution $\Delta E =$ 0.2~eV. An overview on the techniques can be found in Ref.[~\cite{Vondracek2016}]. For metallic samples the Fermi edge $E_{\text F}$ is derived from the kinetic energy at which the $k$-PEEM intensity disappears. 
We define this energy as zero binding energy and expect the error of this Fermi level estimation to be $\pm$ 0.05 eV. $(k_x, k_y)$-cuts at the Fermi surface cover approximately a range $\pm 2$\AA$^{-1}$. \\
STM was done at room temperature under UHV conditions ($p \approx 5\times 10^{−11}$ mbar) using an {\it Omicron VT-STM}.\\
X-ray absorption spectroscopy (XAS) was performed in the total electron yield (TEY) mode at the synchrotron {\it SOLARIS} - {\it XAS} beamline - using circular polarized light at variable temperatures and magnetic fields up to $0.2$~T.\\
Density funtional theory (DFT) calculations were done in the local spin-density approximation (LSDA) using the tight-binding linear muffin-tin orbital method (TB-LMTO) with 2 empty spheres in the basis. As a structure we assume a tetragonal PbO-structure (P4/nmm) $\beta$-FeSe with lattice constants $a = 0.3765$~nm and $c = 0.5518$~nm.

\section{XPS evaluation procedures}
\label{sec:A1}
Fitting was done in the program \texttt{KolXPD} [\url{https://www.kolibrik.net/kolxpd}].
Voigt peaks on a linear background were used for fitting. The estimation of the relative concentrations between Bi 4$d$ and Se 3$d$, in Fig.~\ref{Fig:XPSCEITEC} is based on respective fitting of spin-orbit split Voigt doublet areas, which are normalized to calculated photoionization cross sections~\cite{Yeh1985}. A Shirley background is asumed in all cases. For Fe $2p$ spectra the leading edges are fitted by Donich-Sunjic profiles. Two Voigt components were added to higher binding energies.

\begin{acknowledgments}

We thank prof. Hidde Brongersma for valuable discussions. This work was supported by the Czech Science Foundation Grant
No. P204/19/13659S and EU FET Open RIA Grant No. 766566. The research was partially supported by the Ministry of Education, Youth and Sports of the Czech Republic LQ1601 (CEITEC 2020). Part of the work was carried out with the support of CEITEC Nano Research Infrastructure (MEYS CR, 2016–2019).
We acknowledge CERIC-ERIC support for synchrotron measurements at SOLARIS (project 20182048) and ELETTRA (project 20192095).\\

\end{acknowledgments}



\bibliography{FeCh-literature}

\end{document}